 \documentclass[final,5p,times,twocolumn,authoryear]{elsarticle}

\usepackage{amssymb}
\usepackage{lipsum}

\usepackage{color}
\usepackage{bm}
\usepackage{amsmath}
\newcommand{\Acal}{{\mathcal A}}

\newcommand{\dRM}{{\mathrm d}}
\newcommand{\eRM}{{\mathrm e}}
\newcommand{\iRM}{{\mathrm i}}
\newcommand{\pRM}{{\mathrm p}}

\newcommand{\URM}{{\mathrm U}}

\newcommand{\hphi}{{\hat \varphi}}
\newcommand{\hpsi}{{\hat \psi}}

\newcommand{\LnormalLagr}{ {\mathcal L}}

\newcommand{\ket}[1]{ \left| #1 \right\rangle  }
\newcommand{\bra}[1]{ \langle #1   }
\newcommand{\fmeasure}[1]{ [ \mathrm{D} #1 ] }

\journal{Physica A}

\allowdisplaybreaks
\begin{document}

\begin{frontmatter}



\title{Approximate calculation of functional integrals arising from the operator approach}

\author[first,second,third]{Edik Ayryan}
\ead{ayrjan@jinr.ru}
\affiliation[first]{organization={Meshcheryakov Laboratory of Information Technologies, Joint Institute for Nuclear Research},
            addressline={Joliot-Curie 6}, 
            city={Dubna},
            postcode={141980}, 
            country={Russian Federation}}
\affiliation[second]{organization={State University Dubna, Russian Federation},
            addressline={19 Universitetskaya St}, 
            city={Dubna},
            postcode={141980}, 
            country={Russian Federation}}
\affiliation[third]{organization={A. I. Alikhanyan National Science Laboratory},
            addressline={Alikhanian Brothers Str. 2}, 
            city={Yerevan},
            postcode={0036}, 
            country={Armenia}}
            
\affiliation[fourth]{organization={Bogolyubov Laboratory of Theoretical Physics, Joint Institute for Nuclear Research},
            city={Dubna},
            postcode={141980}, 
            country={Russian Federation}}
 
\affiliation[fifth]{organization={Institute of Experimental Physics, Slovak Academy of Sciences},
            addressline={Watsonova 47}, 
            city={Kosice},
            postcode={ 040 14}, 
            country={Slovakia}}

\affiliation[sixth]{organization={Faculty of Science, \v{S}af\'arik University},
            addressline={Moyzesova 16}, 
            city={Kosice},
            postcode={040 01}, 
            country={Slovakia}}

\author[first]{J\'an Bu\v{s}a}
\ead{busaj@jinr.ru}
\author[fourth,fifth,sixth]{Michal Hnati\v{c}}
\ead{hnatic@saske.sk}
\author[sixth]{Tom\'a\v{s} Lu\v{c}ivjansk\'y}
\ead{tomas.lucivjansky@upjs.sk}

\author[seventh]{Victor Malyutin}
\ead{malyutin@im.bas-net.by}
\affiliation[seventh]{organization={Institute of Mathematics of the National Academy of Sciences of Belarus},
            city={Minsk},
            postcode={220072 }, 
            country={Belarus}}

\begin{abstract}
We apply the operator approach to a stochastic system
belonging to a class of death-birth processes, which
we introduce utilizing the master equation approach. 
By employing Doi-Peliti formalism we recast the master equation in the form of a Schr\"odinger-like equation. Therein appearing
 pseudo-Hamiltonian is conveniently expressed in a suitable Fock space, constructed
 using bosonic-like creation and annihilation operators. The kernel of the associated time evolution operator is rewritten
 using a functional integral, for which
 we propose an approximate method that allows its analytical treatment.
The method is based on the expansion in
eigenfunctions of the Hamiltonian generating given functional
integral. In this manner, we obtain
approximate values for the
probabilities of the system being in the 
 first and second states for the case of the pure birth process.
\end{abstract}

\begin{keyword}
Master equations \sep Schr\"odinger-like equation \sep operator approach \sep functional integrals
\end{keyword}

\end{frontmatter}



\section{\label{sec:intro} Introduction}
Currently, functional integrals constitute a well-established framework of theoretical and mathematical physics \citep{schulman2012,kleinert_book}. Not only they
provide convenient language in which many physical problems can be formulated, but also offers non-trivial methods for their 
solutions \citep{Zinn2007}.
Although, their origins can be traced back to quantum physics \citep{feynman2010quantum}, their widespread applications 
cover diverse research fields such as high-energy physics, statistical physics, critical phenomena, non-equilibrium physics, econophysics, etc., \citep{Zinn2007,kardar_book,kleinert_book}. 
 The usefulness of functional integrals
in theoretical physics can be discussed at various levels. 
First, they provide a non-trivial framework in which physical problems can be formulated. For example, quantization of non-abelian gauge theories is almost impossible to tackle without the use of functional integrals \citep{Zinn2002,Faddeev18}.

Second, functional integrals offer crucial insight into the mathematical relations between
quantum field theory and statistical physics of
continuous phase transitions \citep{Zinn2007}. 
The latter can be thought of an imaginary in-time version of the former.

Third, formulation of physical theory in
terms of functional integrals often allows a
use of sophisticated calculational methods
such as instantons, renormalization group, operator product expansion, etc.
Results obtained by these means provides a complementary route to assess strongly non-linear
regimes. Quite often they are in very good
agreement with other theoretical or numerical approaches (such as Monte Carlo simulations). 
For example, in model A of critical dynamics \citep{adzhemyan2022}
the recently obtained five-loop calculation using field-theoretic approach
based on a functional formulation agrees well with the numerical methods.
Further, in reaction-diffusion problems \citep{Lee1994,Tauber2014}, functional methods augmented with renormalization group calculations provides a robust
substantiation of scaling regimes accompanied with perturbation techniques 
for universal quantities such as various critical exponents or amplitudes.

From a strict mathematical perspective, the associated functional measure is not a measure in the mathematically strict sense. 
As a consequence many different
approaches to Feynman integrals have been suggested, 
different ways for their construction were developed and last but not least appropriate methods for their analysis were 
offered~\citep{morette2000}.
For practical purposes, it is of utmost importance
 to possess a reliable method for the evaluation of a functional integral. As actual calculations show this often presents a
 rather formidable task \citep{kleinert_book,grosche2014handbook}.
Since the functional integrals find ever wider applications, it is of utmost importance to have effective calculational methods at one's disposal for their evaluation. 
  In practice, most functional integrals are difficult, if not impossible, to be exactly solvable. One of the possible
 routes to undertake lies in an approximate treatment of functional integrals and in the generalization of existing methods to tackle new
types of functional integrals. 
In the last fifty years, the widespread occurrence of functional integrals stimulated the development of various numerical methods for their approximate calculation and led to
  a significant progress in this regard.
 Nowadays, several promising methods exist for computing different types of functional integrals \citep{Hnatic2023,grosche2014handbook}.
 As discussed in detail in \citep{Hnatic2023}, functional integral formulation analyzed by means of eigenfunction
expansion of the associated Hamiltonian is particularly effective in an analysis of large-time asymptotics of a given physical model. 

One of the most prominent areas, in which functional integrals find their applications, is
statistical physics, especially the field of critical phenomena and
phase transitions \citep{Zinn2007,Vasilev2004,Tauber2014}.  
The models are often phenomenologically modeled through the Langevin equation. This mathematically represents a stochastic differential equation with prescribed properties of a random force term. Quite often
such an approach is based on certain heuristic considerations
and as such can not be considered microscopic.
However, in the realm of non-equilibrium physics there
exists a vast set of problems related to
 interacting (classical) stochastic particle systems \citep{Tauber2014,HHL16}
 for which it is possible
 to construct microscopically well-founded models. What these models share is the property that their
configurational dynamics can be captured by a discrete set of states, whose time evolution is governed by the master equation \citep{Kampen}. Areas of application include 
various reaction-diffusion models
\citep{Kamenev2004,Kamenev2006,Tauber2005,krapivsky2010kinetic}, population
dynamics models \citep{Meerson2006}, neural network models
 \citep{bressloff2009,bressloff2013}, etc.
 
In this regard, the so-called Doi-Peliti approach is a particularly effective theoretical approach~\citep{Doi1976a,Doi1976b,Peliti1985,cardy2008non} for analytical treatment of classical systems with a non-conserved number of interacting degrees of freedom. 
It exploits a formal analogy between the master equation and Schr\"odinger equation, which allows a concise representation of the formal solution in terms of a
functional integral with a specific Lagrangian function.
The Doi-Peliti approach thus
provides a direct transition from the original model to the
 field-theoretical model (represented
by a suitable effective action) without direct use of the
Langevin equation, which is utilized in the combinatorial approaches \citep{Gardiner,Kampen}.
 We note that this is not just theoretical pedantry. 
As observed by one of the authors in the past
\citep{AHHN1998}, the latter approach actually yields a wrong effective model 
for the self-interacting scalar field (corresponding to the 
binary annihilation process).
On the other hand, the master equation approach rewritten in terms
of functional integrals leads to a correct effective model \citep{Tauber2014}.
The ensuing model might display the presence of a non-trivial scaling regime, which is amenable to 
sophisticated field-theoretic
methods such as renormalization group and accompanying techniques~\citep{Tauber2005,Tauber2014,HHL16}. 
In this way, we may gain invaluable insights into the asymptotic behavior of a model \citep{Dickman2003,klauder2010modern}.
Paradigmatic models of such reaction systems
encompass pair annihilation process
 \citep{Peliti1986,Lee1994,Hnatic2011b,Hnatic2013}, certain
one-step processes \citep{Hnatic2016}, the decay process and the
Malthus-Verhulst process \citep{Dickman2003}, hybrid systems involving discrete and continuous stochastic processes \citep{Bressloff2021}, branching processes \citep{pruessner2018}, two-species binary annihilation process \citep{Lee1995,cardy2008non}, trapping reaction \citep{Rajesh2004,VollLee2018,VollLee2020} and many others.

Following the previous work of some of the present authors \citep{Hnatic2023}, we adopt
 the operator approach (based on Doi-Peliti formalism) to a general one-step model. We rewrite the
kernel of the evolution operator through a functional
integral, in which integration is carried out over coordinates and
momenta. Once integration over momenta is performed, we obtain the
 modified functional integral, in which 
 integration is carried out only over
coordinate variables. Then we express the generating function and the probabilities of being in the $n$-th state through a functional integral amenable to approximate treatment.
A representation in terms of a functional integral allows us
to expand
the number of methods for studying stochastic systems described by
the master equation. In particular, various approximate methods
for calculating functional integrals can be employed. 
Here, for the functional integral arising in the operator approach, we propose a method for its approximate calculation based on the expansion in the eigenfunctions of the Hamiltonian that generates the
functional integral \citep{risken1984}. Utilizing this method for the pure birth process, we obtain the approximate values of the probabilities of being in the $n$-th state for $n=0$ and $n=1$, respectively.

This paper is organized as follows. In Sec.~\ref{sec:transition} the model is formulated and the main elements of the theoretical approach based on Doi--Peliti formalism are presented.  In Sec.~\ref{sec:specific} we
 restrict ourselves to a particular model that allows analytical treatment.
An approximate evaluation of the arising functional integral is carried 
out in Sec.~\ref{sec:calculation}.  Sec. \ref{sec:conclusion} is reserved for concluding remarks. 

 {\section[Functional integral representation of the master equation]{Functional integral representation of the master equation
\phantom{II.~~~}}
\label{sec:transition}
}
In this work, we are interested in the statistical properties
of a specific one-step process \citep{Kampen,Gardiner} whose range 
consists of
non-negative integers $n=0,1,2,\ldots$ The integer 
 $n$ effectively enumerates permissible discrete states in which a system can occur.
The relevant statistical
 quantity is $\pRM_n = \pRM_n(t)$ being the probability
 for a system to be in $n$-th state at time $t$. The
transition matrix between different states takes the following form
\begin{equation}
    W_{n \rightarrow m} = \lambda_n \delta_{m,n+1} + \nu_n\delta_{m,n-1},
\end{equation}
where 
 $\delta_{i,j}$ is the Kronecker symbol ($\delta_{i,j}= 1$ if $i=j$ and zero otherwise), $\lambda_n$ is the probability per unit time for a transition from $n$-th state to $(n+1)$-th state, whereas $\nu_n$ is the probability
per unit time to jump in the opposite direction, i.e., from $n$-th state 
to the $(n-1)$-th. We can interpret 
parameters $\lambda _{n} $ and $ \nu _{n}$  as the infinitesimal intensities for the birth and death process, respectively.
\begin{figure}
    \centering
    \includegraphics[width=0.5\textwidth]{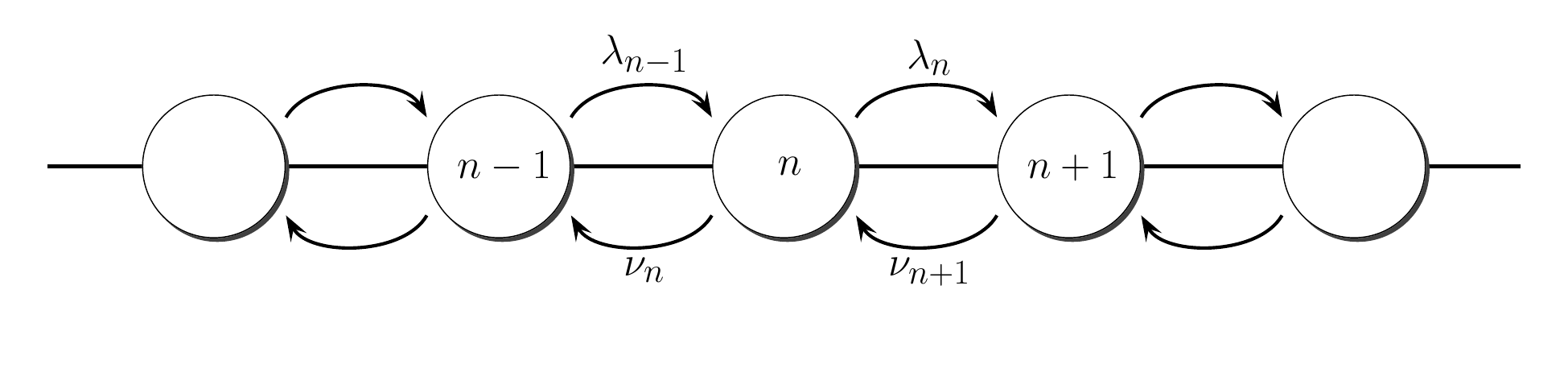}
    \caption{Graphical representation of the studied one-step process with its transition rates.}
    \label{fig:poisson}
\end{figure}
 The process is schematically depicted in Fig.~\ref{fig:poisson}.
The time evolution of the process is governed by the following
  master equation 
 \begin{equation}
 \frac{\dRM {{\pRM}_{n}}(t)}{\dRM t}
 =\lambda
_{n-1}{{\pRM}_{n-1}}(t)+\nu
_{n+1}{{\pRM}_{n+1}}(t)-(\lambda _{n}+\nu
_{n}){{\pRM}_{n}}(t), 
 \label{eq:basic}
\end{equation}
which is valid for any $n\ge 1$.
To finalize the setup, let us note that the state $n=0$ 
fulfills the following equation
\begin{equation}
  \frac{\dRM {{\pRM}_{0}}(t)}{\dRM t}
  = \nu_1 \pRM_1(t) - \lambda_0 \pRM_0(t).
  \label{eq:basic2}
\end{equation}
 
Probabilities $\pRM_n$ are necessarily restricted by two 
obvious conditions
\begin{equation}
   \pRM_n(t)\ge 0, \qquad \sum\limits_{n=0}^{\infty } \pRM_n(t) = 1, 
   \label{eq:conditions}
\end{equation}
where the latter corresponds to the normalization condition.
   Further, we consider the specific case of general one-step
   processes~\ref{eq:basic} corresponding to the
   choice 
\begin{equation}
   \lambda _{n}=\mu _{1}(n + N_0), \quad \nu _{n}=\mu _{2}n. 
   \label{eq:model_param}
\end{equation}
   Here,  $N_0$ denotes the number of members in the initial population 
   size at the time $t=0$.
   Thus for $n\ge 1$ we have to deal with the following set of equations
\begin{align}
\frac{\dRM {\pRM}_{n}(t)}{\dRM t}
 & = 
{\mu}_{1}(n - 1 + N_0){\pRM}_{n-1}(t)+{{\mu
}_{2}}(n+1){\pRM}_{n+1}(t) 
\nonumber\\ 
& - 
\left(
 \mu _{1} ( n + N_0 ) +{ \mu_{2}}n
\right) \pRM_{n}(t).
\label{eq:master}
\end{align}
The probability $\pRM_0$ satisfies the corresponding version 
of Eq.~(\ref{eq:basic2}). The introduced process described by Eq.~(\ref{eq:master}) is also known in the literature as a process with linear growth.
For specific value $\mu_2=0 $  this process corresponds to the
so-called  Yule process (a pure birth process) \citep{newman2005,ross2007}.

From a theoretical point of view, the main interest lies in an investigation of various statistical quantities such as moments 
$M_k$ or correlations functions as a function of time. 
The $k-$th moment is standardly defined \citep{Kampen} as 
\begin{equation}
   M_k(t) = \overline{n^k} = \sum_{n=0}^\infty n^k \pRM_n(t),
   \label{eq:moments}
\end{equation}
for integer $k \ge 1$, and a bar denotes statistical averaging. For instance, $M_1(t)$ corresponds to the mean particle number at time $t$, etc. 

Although, the techniques and methods we employ in this paper are well-covered in the literature
\citep{Peliti1985,Dickman2003,cardy2008non,Tauber2014,HHL16}, we summarize the main points of the used formalism, thus covering necessary background information on the use of path (functional) integrals 
for stochastic processes.
 Our starting point is to recast the master 
 equation~(\ref{eq:master}) into a suitable functional integral form, which can be further analyzed by approximate methods.
To this end, first, following the works \citep{Dickman2003,Tauber2005,Hnatic2013,Hnatic2016}, we convert the master equation to a Schr\"odinger-like equation 
containing specific Hamiltonian that effectively encodes reaction
rates of a given stochastic process. 
 The associated state space
 closely resembles Fock space in quantum mechanics \citep{grassberger1980,cardy2008non}.
To perform the required mapping,
we assign to a state $n$ a ket vector
 $\ket{n}$ of the Hilbert space, which is composed of a vacuum state
 $\ket{ 0 }$ with zero "particles" (excitations), along with $n$ particles created by a successive application of the creation operator $\pi$, i.e.,
 \begin{equation}
    \ket{ n } = \pi^n \ket{ 0 }.
    \label{eq:n_ket_state}
 \end{equation}
  The creation operator $\pi$ and the annihilation operator $a$ form
 bosonic-like ladder algebra and obey the rules \citep{grassberger1980,Peliti1985,Dickman2003}
\begin{equation}
  \pi \left| n \right\rangle =\left| n+1 \right\rangle,
  \quad a\left| n \right\rangle =n\left| n-1 \right\rangle.
  \label{eq:api_definition}
\end{equation}
The vacuum state $\ket{0}$ is subject to standard condition $a \ket{0} = 0$.
Also note that the numerical factors in Eq.~(\ref{eq:api_definition}) are different from the standard
rules for quantum mechanical oscillator \citep{schwabl2007quantum}.
From relations (\ref{eq:api_definition}) we directly infer the  commutation relation between the creation and annihilation operator
that takes a standard form
\begin{equation}
    [a,\pi] = a\pi - \pi a = 1.
    \label{eq:commutation}
\end{equation}
  The Hilbert space is further equipped with the inner product between two states 
 $\bra{m} \ket{n} = n!\delta_{n,m}.$ Among others this ensures that  
 the identity can be written as 
 $1 = \sum_n \frac{1}{n!} \ket{n} \bra{n} | $ and also that operators $a$ and $\pi$ are Hermitian conjugate to each other \citep{grassberger1980,Peliti1985}.

 The whole set of probabilities
$\{ \pRM_n(t) \}_{n=0}^\infty$ is then
effectively encoded in  the state vector 
$\ket{ \varphi (t) }$ introduced as
\begin{equation}
 \ket{ \varphi (t) }
=\sum\limits_{n=0}^{\infty }{{{\pRM}_{n}}(t)
 \ket{ n } }.    
 \label{eq:ket1}
\end{equation}
 Note that such a definition
 leads to substantial differences with usual relations from quantum mechanics
  \citep{cardy2008non}. In particular, physically relevant quantities can not be
 given by bilinear expressions of ket vector $\ket{\varphi(t)}$.
 
 For the introduced state (\ref{eq:ket1}) we can readily derive a dynamic equation 
\begin{align}
  \frac{\dRM \ket{ \varphi (t) } }{\dRM t}
 & = \sum\limits_{n=0}^{\infty
}
\frac{\dRM {\pRM}_{n}(t)}{\dRM t}
\left| n \right\rangle  \nonumber\\
& = 
\sum\limits_{n=0}^{\infty } [
\mu_1(n - 1 + N_0)\pRM_{n-1}(t)+
\mu_2 (n+1) 
\pRM_{n+1}(t)
\nonumber \\
& - 
(\mu_1 [n + N_0] + \mu_2 n) \pRM_n(t)
] \ket{ n }  .
\label{eq:ket2}
\end{align}
To rewrite the latter sum we need the following auxiliary relations
\begin{align} 
  n  \ket{ n }  & =  \pi a \ket{ n }, 
  \label{eq:aux1}
  \\
  (n-1+N_0)\ket{ n }  & = (\pi a\pi
+(N_0 - 1)\pi ) \ket{ n - 1 },
\label{eq:aux2}
\\
(n+1) \ket{ n }  & = a \ket{ n + 1 } ,
\label{eq:aux3}
\\
(n+N_0) \ket{n}  & =  (a\pi + N_0 - 1) \ket{n},
\label{eq:aux4}
\end{align}
which can be derived straightforwardly using the commutation relation~(\ref{eq:api_definition}).
Employing them we readily rewrite Eq.~(\ref{eq:ket2}) as follows
\begin{align}
\frac{\dRM \ket{ \varphi (t) } }{\dRM t}
& =  \mu_1(\pi a\pi +( N_0 -1)\pi)
\sum\limits_{n=0}^{\infty }\pRM_{n}(t)
 \ket{ n } \nonumber\\
& + \mu_2 a\sum\limits_{n=0}^{\infty}
 \pRM_{n}(t) \ket{ n } -\mu_1
 (a\pi + N_0 - 1) \nonumber\\ 
 &\times
  \sum\limits_{n=0}^{\infty }
 \pRM_{n}(t) \ket{ n }
  -  \mu_2 \pi a
 \sum\limits_{n=0}^{\infty} \pRM_{n}(t) 
 \ket{ n } .
 \label{eq:dynamic_eq0}
\end{align}
Let us note that in deriving the expression on the right-hand side we have used the obvious fact $a \ket{0} = 0$ to obtain indicated lower limits of the sums.
Further we observe that Eq.~(\ref{eq:dynamic_eq0}) takes a form of an imaginary-time Schr\"odinger-like equation 
\begin{equation}
  \frac{\dRM \ket{  \varphi (t) }  }{\dRM t}
  = L \ket{ \varphi (t) },
 \label{eq:dynamic_eq}
\end{equation}
where we have introduced the Liouville operator $L$
 (also known in the literature as Liouvillian \citep{Peliti1985} or pseudo-Hamiltonian \citep{Tauber2014}) 
\begin{equation}
  L \equiv \mu_1(\pi
a\pi +(N_0 - 1)\pi )+\mu_2a-\mu_1(a\pi +N_0-1)-\mu_2\pi a.
  \label{eq:liouville}
\end{equation}
Formal exponentiation of Eq.~(\ref{eq:dynamic_eq})
 yields the solution
\begin{equation}
  \ket{ \varphi(t) } = \URM_t \ket{ \varphi_0 }, \qquad 
  \URM_t = \exp\left( L t \right),
  \label{eq:formal_solution} 
\end{equation}
where $\URM_t$ plays the role of time evolution operator akin to
quantum mechanics and $\ket{\varphi_0 } = \ket{\varphi(t=0)}$ is a  state
specified at initial time instant $t=0$.

In dealing with operator expressions, it is
advantageous to work with the normal ordered
form of a given operator $A$, which is obtained
by commuting all creation operators to the left of all annihilation operators \citep{Vasilev1998}. 
An operator $A$ written in its 
 its normal ordered form  thus can be
  formally represented as a sum
\begin{equation}
 A = \sum_{m,n}\Acal_{m,n} \pi^m a^n, 
 \label{eq:normal_operator}   
\end{equation}
where $\Acal_{m,n}$ are complex coefficients.
Further, it is convenient to rewrite the Liouville operator $L$ (\ref{eq:liouville}) in its normal ordered form. This can be readily achieved
 using commutation relation (\ref{eq:commutation})
 yielding
\begin{align}
  L & = & \mu_1 \pi(\pi a + N_0 )  + \mu_2 a 
  -\mu_1(\pi a + N_0) - \mu_2 \pi a 
  \label{eq:liouville2a}
  \\
  & = & (\pi - 1)\left[
  \mu_1 ( \pi a + N_0) - \mu_2 a
  )
  \right]
   .
  \label{eq:liouville2b}
\end{align}
For the forthcoming use of functional integrals, we
invoke Bargmann-Segal space 
\citep{bargmann1961,Peliti1985,schulman2012,itzykson2012quantum}, which is based
on an equivalence between analytic functions $\hphi(z)$ of a complex
variable $z$ with the Fock states $\ket{\varphi}$
 from the aforementioned Hilbert space formed
 by linear combinations of
 states~(\ref{eq:n_ket_state}). 
The relation can be succinctly stated as
\begin{equation}
    \ket{\varphi} = \sum_n \varphi_n \ket{n} \longleftrightarrow 
    \hphi(z) = \sum_n \varphi_n z^n ,
    \label{eq:bargmann}
\end{equation}
where $\varphi_n = \frac{1}{n!}\bra{n}\ket{\varphi}$.
The scalar product between two states 
$\ket{\varphi}$ and $\ket{\psi}$
is then expressed through
an integral formula
 $ \bra{\varphi} \ket{\psi} = \int 
 \frac{\dRM x \dRM x'}{2\pi}
   \hphi(x) \hpsi(\iRM x') \exp \left(-\iRM xx'\right) $, where both integrals are taken over
   the entire real axis.
To a given operator $A$ with matrix elements $A_{m,n} = 
\bra{ m } | A \ket{ n }$ we associate
the kernel function
\begin{equation}
  A(z,\xi) = \sum_{m,n} \frac{z^m}{m!} \frac{\xi^n}{n!}
  A_{m,n}.
  \label{eq:kernel}
\end{equation}
For our discussion, two important formulas are needed \citep{Peliti1985,Dickman2003}.
First, if there are two states $\ket{\psi}$ and $\ket{\varphi}$ related by $\ket{\psi} = A\ket{\varphi}$, then we have a relation for the function $\hpsi(z)$ associated with the state $\ket{\psi}$
\begin{equation}
  \hpsi(z) = \int \frac{\dRM x \dRM x'}{2\pi} 
  A(z,x)
  \hphi(\iRM x') \exp \left(
  -\iRM x x'
  \right).
  \label{eq:kernel_states}
\end{equation}
The second formula is related to a kernel of the product of two operators 
$A$ and $B$, and can be presented in the following form~\citep{Peliti1985}
\begin{equation}
    AB(z,\xi) = \int\frac{\dRM x \dRM x'}{2\pi}
    A(z,x) B(\iRM x',\xi)
    \eRM^{-\iRM x x'}.
    \label{eq:kernel_product}
\end{equation}
In this formula both integration variables $x$ and $x'$ are real. However, it is also possible to 
 transfer complex unity on the corresponding 
 $x'$ variable as is commonly done, see e.g., \citep{Janssen92,Tauber2014}.
The kernel associated with the normal order form
of the 
operator $A$ is  introduced as follows
$\Acal(z,\xi) = \sum_{m,n} \Acal_{m,n} z^m \xi^n$, which
 assumes slightly different normalization than the previously defined
 kernel~(\ref{eq:kernel}). The coefficients
  $\Acal_{m,n}$ have been introduced previously in Eq.~(\ref{eq:normal_operator}).
It can be shown \citep{itzykson2012quantum} that kernels $A(z,\xi)$ and $\Acal(z,\xi)$ are proportional to each other through
the simple relation $A (z,\xi) = \eRM^{z \xi} \Acal(z,\xi) $.

With this machinery at our disposal, 
by relatively straightforward algebraic manipulations \citep{Peliti1985,Dickman2003}
we find a functional representation for the time evolution operator $\URM_t$ introduced in Eq.~(\ref{eq:formal_solution}). The derivation
 consists essentially of two steps.
First, with the operator $\URM_t$ we associate the
corresponding kernel $\URM_t(z,\xi)$ using the definition (\ref{eq:kernel}).
Second, employing the well-known
Trotter product formula \citep{schulman2012} according to which one
may formally express $\URM_t$ as the infinite product
\begin{equation}
    \URM_t = \lim_{n\rightarrow\infty} 
    \left(
    1 + \frac{t}{n} L
    \right)^{n}
    = 
    \lim_{\tau \rightarrow 0} 
    \left(
    1 + \tau L
    \right)^{t/\tau}
    ,
    \label{eq:trotter}
\end{equation}
where in the last relation we have introduced for future convenience a small time interval $\tau$ defined as 
\begin{equation}
    \tau \equiv \frac{t}{n}.
    \label{eq:def_tau}
\end{equation}
Then the kernel function $L$ associated with each of the $n$ factors in the Trotter formula~\eqref{eq:trotter} we rewrite through its normal ordered kernel function $\LnormalLagr$ as follows
\begin{equation}
    \left(1 + \tau L (z,\xi)\right)  = 
    \eRM^{z\xi} \left(1 + \tau \LnormalLagr(z,\xi) \right). 
\end{equation}
Applying this with the formula for the product of two kernels (\ref{eq:kernel_product}) we finally arrive at the discretized expression for the kernel of the time evolution operator
\begin{align}
\URM_t(z,\xi) & =  \lim\limits_{n\to
\infty}\int\prod\limits_{j=1}^{n-1}
\frac{\dRM\psi_j \dRM\psi'_j}{2\pi} \exp
\biggl( -\sum\limits_{k=1}^{n}\biggl[ \iRM\psi'_{k}(\psi_k-\psi_{k-1})
\nonumber\\
& - 
\tau
\LnormalLagr(\iRM\psi'_{k},\psi_{k-1}) \biggl]
+z\psi_n \biggl),
\label{eq:Utz_discrete1}
\end{align}
where the standard approximation for the exponential function has been employed. To comply with the usual field-theoretic notation
with have denoted the intermediate points by $\psi_j$ and $\psi_j'$.
The expression in the brackets of Eq.~(\ref{eq:Utz_discrete1}) takes a slightly different form than
 that common in the literature \citep{Peliti1985,Dickman2003}. 
 The discretization points were chosen as 
$\psi'_{k}=\psi'\left(t_k\right)$,
$\psi_{k}=\psi(t_k)$, where $t_k = k\tau $ and
$k=0,1,\ldots,n$. The appropriate boundary conditions are
$\psi_{0} = \xi$, and $\iRM\psi'_{n} = z$.

Since the expression (\ref{eq:Utz_discrete1})
 contains an infinite number of intermediate
points it is natural to express it formally via  
the functional integral \citep{Peliti1985,schulman2012}
\begin{align}
\URM_t(z,\xi) & =  \int
\fmeasure{ \psi } \fmeasure{\psi^\prime}
\exp\biggl(
-\int\limits_0^t \dRM s\, 
[\iRM\psi'(s)\dot{\psi}(s)
\nonumber\\
& - 
\LnormalLagr(\iRM\psi'(s),\psi(s))] + z\psi(t)
\biggl)
\label{eq:Utz},
\end{align}
where $\fmeasure{\cdots}$ is the measure in the functional space, and hereinafter
 the dot symbol indicates the time derivative.
 Let us remark that in (multi-dimensional) field theories, variable $\psi'$  corresponds to the (complex-valued)
 Martin--Siggia--Rose response
field \citep{MSR1973,Vasilev2004,Zinn2007,Tauber2014}.
Let us stress that continuous formulas for functional integrals such as~(\ref{eq:Utz}) are inherently ambiguous \citep{wissel1979}, and mathematically consistent definition can be given only by explicit specification of its discretized form. 

For the kernel function (assumed at a given instant)
pertaining to the normal
 ordered Lioville operator~ (\ref{eq:liouville2a}), the needed relation simply amounts \citep{Peliti1985,Dickman2003} to the replacement $\pi \rightarrow \iRM\psi'$ and $a \rightarrow \psi$ in the corresponding expression. We thus arrive at
\begin{align}
  \LnormalLagr (\iRM\psi',\psi) 
 & =  \mu_1((\iRM\psi')^2\psi+
N_0\iRM\psi')+\mu_2\psi-\mu_1(\iRM\psi'\psi+N_0)
\nonumber\\
& -  \mu_2\iRM\psi'\psi.
\label{eq:normal_lagr}
\end{align}

Let us return to Eq.~(\ref{eq:Utz_discrete1}) and analyze it further.
 To obtain a closed-form expression for
 the kernel $\URM_t(z,\xi)$ it is advantageous
 to introduce the function $\hat{\URM}_t(\psi_n,\psi_0)$ defined as
\begin{equation}
  \hat{\URM}_t(\psi_n,\psi_0)  \equiv \int\frac{\dRM\psi'_n}{2\pi}
  \URM_t(z,\psi_0)\exp
  \left(
    -z\psi_n
  \right).
  \label{eq:Utz_discrete2}  
\end{equation}
Using previous relations this function corresponds to the
following expression
\begin{equation}
\int\frac{\dRM\psi'_n}{2\pi}
  \URM_t(\iRM \psi_n',\psi_0)\exp
  \left(    - \iRM \psi_n' \psi_n
  \right)
\end{equation}  
whose discretized version takes the form
\begin{align}  
  &
  \lim  \limits_{n\to
  \infty}
  \int  
  \prod\limits_{j=1}^{n-1}\frac{\dRM \psi_j \dRM \psi'_j}{2\pi}
  \frac{\dRM \psi'_n}{2\pi} 
  \exp
  \biggl(
  -\sum\limits_{k=1}^{n} [
  \iRM\psi'_k(\psi_k-\psi_{k-1})
  \nonumber\\
  & -
  \tau
  \LnormalLagr(\iRM\psi'_k,\psi_{k-1})
  ]
  \biggl).
  \label{eq:Utz_discrete3}
\end{align}
 Recall that in the previous
 formula~(\ref{eq:Utz_discrete1}) the expression
 in the argument of the exponential function
 depends on variables
$\psi_0,\ldots,\psi_{n-1}$ and $\psi'_1,\ldots,\psi'_n$, respectively. 
Therefore, after the integration over variables
$\prod\limits_{j=1}^{n-1}\dRM\psi_j \dRM\psi'_j$ explicit dependence
on values  $\xi =\psi_0$, $z=\iRM\psi'_n$, and $\psi_n$ remains.
On the other hand, in 
expression~(\ref{eq:Utz_discrete3}), additional integration ensures that the exponent depends only on the values
$\xi =\psi_0$ and $\psi_{n}$, respectively.

 Inserting the expression (\ref{eq:normal_lagr}) into Eq.~(\ref{eq:Utz_discrete3}) for the
 Liouville operator
$\LnormalLagr(\iRM\psi'(s),\psi(s))$ yields
\begin{align}
 \hat{\URM}_t(\psi_n,\psi_0) &= \lim\limits_{n\to
\infty}\int\prod\limits_{j=1}^{n-1} \dRM\psi_j
\prod\limits_{l=1}^{n}\frac{\dRM\psi'_l}{2\pi}
\exp
\biggl(
-\sum\limits_{k=1}^{n} \biggl[ \iRM\psi'_k
\nonumber\\
&\times
(\psi_k-\psi_{k-1})+ 
\tau
\mu_1 (\psi'_k)^2\psi_{k-1}
\nonumber\\
 & +   
\tau
\iRM\psi'_{k}
(\mu_1\psi_{k-1}+\mu_2\psi_{k-1}-\mu_1 N_0)
\nonumber\\
& + 
\tau
(\mu_1 N_0 - \mu_2\psi_{k-1}) \biggl]
\biggl).
\label{eq:Utz_discrete4}
\end{align}
 In this functional integral, integration is carried out over
 the real variables $\{\psi_j\}_{j=1}^{n-1}$ and 
 $\{\psi'_j\}_{j=1}^n$. In principle, such procedure might be interpreted as an integration over generalized coordinates and momenta. Note that integration over the response variables
 $\prod\limits_{l=1}^{n}\dRM\psi'_l$ can be performed straightforwardly.
  To this end, let us recall an auxiliary formula involving the elementary Gaussian integral
\begin{align} 
 &   \int
\dRM\psi'_k \exp
\biggl(
-
\tau
\mu_1(\psi'_{k})^2\psi_{k-1} -\iRM\psi'_{k}(\psi_{k}-
\psi_{k-1})
\nonumber \\
& - \iRM\tau\psi'_{k}(\mu_1\psi_{k-1}+\mu_2\psi_{k-1}
-\mu_1 N_0)
\biggl)
 =   
\sqrt{\frac{\pi}{\tau \mu_1\psi_{k-1}}}
\nonumber\\
 & \times 
 \exp
\left(
-
\frac{ \left[
 \psi_k-\psi_{k-1}+
 \tau
 ( ( \mu_1+ \mu_2)\psi_{k-1}
-\mu_1 N_0)
\right]^2}{4\tau\mu_1\psi_{k-1}}
\right),
\label{eq:Utz_discrete5}
\end{align}
which we have written appropriately for our purposes.
Employing this formula in Eq.~(\ref{eq:Utz_discrete3})  the expression for $\hat{\URM}_t(\psi_n,\psi_0)$ becomes
\begin{align} 
   \hat{\URM}_t(\psi_n,\psi_0) & =
   \exp (-\mu_1 t N_0)
\lim\limits_{n\to
\infty}\int\prod\limits_{j=1}^{n-1}\dRM \psi_j
 \prod\limits_{l=1}^{n}
\frac{1}{2\sqrt{\pi \tau \mu_1\psi_{l-1}  }}
\nonumber\\
 & \times
\exp
\biggl(
-
\sum\limits_{k=1}^{n}
\frac{1}{4 \tau\mu_1\psi_{k-1} }
\biggl[
(\psi_k-\psi_{k-1}+
\tau
( (\mu_1+
\mu_2)\psi_{k-1}
\nonumber\\
& -\mu_1 N_0)
)^2-4
\tau^2 \mu_1\mu_2\psi_{k-1}^2
\biggl]
\biggl).
\label{eq:Utz_discrete6}
\end{align}
{\section{Specific model $\mu_2 = 0$} \label{sec:specific}}
 The final formula~(\ref{eq:Utz_discrete6}) 
  from the previous Sec.~\ref{sec:transition}
 appears rather formidable.
 To the best of our knowledge, in its full generality, it can be analyzed only
 by numerical means. Related technical obstacles can be traced down
 to the presence of the last term 
 $4\tau^2\mu_1\mu_2\psi_{k-1}^2$ in the numerator of Eq.~(\ref{eq:Utz_discrete6}).
 However, for vanishing parameter $\mu_2$ it is possible
 to analyze it analytically. Therefore, in what follows
 we concentrate on the special case  $\mu_2=0$. Then the function $\hat{\URM}_t(\psi_n,\psi_0)$
   considerably simplifies   
\begin{align}
    \hat{\URM}_t(\psi_n,\psi_0)& =\exp \left( -\mu_1 t N_0 \right)
     \lim\limits_{n\to
    \infty}\int\prod\limits_{j=1}^{n-1}\dRM \psi_j
   \prod\limits_{l=1}^{n}
   \frac{1}{2\sqrt{\pi \tau \mu_1\psi_{l-1} }}
   \nonumber\\
   &\times
\exp
\left(
- 
\sum\limits_{k=1}^{n}\frac{
\left[
\psi_k-\psi_{k-1}+
\tau
(\mu_1\psi_{k-1}-
\mu_1 N_0)
\right]^2}{4\tau\mu_1\psi_{k-1}}
\right).
\label{eq:Ut_simple}
\end{align}
The crucial observation for the ensuing analysis 
consists in an identification of this formula with a transition probability density function
  (TPDF) for a particular type of stochastic differential equation (SDE)
  \citep{langouche2010,wio2013,bennati1999}.
  By direct inspection, we observe that the stochastic process
  corresponding to Eq.~(\ref{eq:Ut_simple}) obeys the following It\^o SDE
\begin{equation}
  \dRM\psi =\mu_1(N_0 - \psi )\dRM t + \sqrt{2\mu_1\psi}\,\dRM W,
  \label{eq:SDE}
\end{equation}
where now $\psi = \psi(t)$ is an auxiliary stochastic variable, and
$\dRM W = \dRM W(t)$  is an increment of the Wiener process \citep{Gardiner}.

For convenience, we replace the variable $\psi$ with the new variable
$y$ through the substitution
\begin{equation}
  \psi =\frac{\mu_1}{2}y^2 .
  \label{eq:psi_y}
\end{equation}
In order to derive SDE for $y$ let us recall the well-known  It\^o's formula \citep{Gardiner}, which can be stated in the form
\begin{align}
   \dRM f[x(t)] & =  \left(
   a[x(t),t] f'[x(t)] +\frac{1}{2}b[x(t),t]^2 f''[x(t)]
   \right)
   \dRM t
   \nonumber \\
   & +  b[x(t),t] f'[x(t)]\,\dRM W(t), 
   \label{eq:ito_formula}
\end{align}
where the stochastic variable $x$ is subject to SDE
of the form
$\dRM x = a[x(t),t]\dRM t + b[x(t),t]\,\dRM W(t)$ and
$f[x(t)]$ is any twice differentiable function of its argument.
From Eqs.~(\ref{eq:SDE})--(\ref{eq:ito_formula}) we readily obtain
 the appropriate SDE for the variable $y$ 
 \begin{equation}
  \dRM y=\left(\frac{1}{y}
  \left(N_0-\frac{1}{2}\right)-\frac{\mu_1y}{2}\right)
  \dRM t + \dRM W.
  \label{eq:stoch_eq_y}
\end{equation}
We now proceed in the opposite direction and return back
 transform SDE (\ref{eq:stoch_eq_y}) to its functional integral representation \citep{langouche2010,wio2013,bennati1999}.
Hence, for the function
$\hat{\URM}_t\left( \mu_1y_{n}^{2}/2, \mu_1 y_{0}^{2}/2\right)$
we find 
\begin{align} 
\hat{\URM}_t\left(\dfrac{\mu_1}{2}y_{n}^{2},\dfrac{\mu_1}{2}y_{0}^{2}\right) &=
\exp
\left(
  -\mu_1 t N_0
\right) \int 
 \fmeasure{y}  
\exp
\biggl(
-\frac{1}{2}\int\limits_{0}^{t}\dRM s\,
\biggl(\dot{y}(s)
\nonumber\\
& -\frac{N_0-\frac{1}{2}}{y(s)}+\frac{\mu_1 y(s)}{2}\biggl)^2 
\biggl)
.
\label{eq:func_integral}
\end{align}
Short introspection reveals that the corresponding discretized form
 of this expression reads
\begin{align} 
\hat{\URM}_t\left(\dfrac{\mu_1}{2}y_{n}^{2},\dfrac{\mu_1}{2}y_{0}^{2}\right) &=
\exp\left(
 -\mu_1 t N_0
\right)
\lim\limits_{n\to
\infty}\int\prod\limits_{j=1}^{n-1}
 \dRM y_j\prod\limits_{j=1}^{n}\frac{1}{\sqrt{2\pi \tau }}
 \nonumber\\
 &\exp \biggl(
-\frac{1}{2\tau}
\sum\limits_{k=1}^{n}\biggl[(y_k-y_{k-1})
-\tau \biggl(\frac{N_0-\frac{1}{2}}{y_{k-1}}
\nonumber\\
& -
\frac{\mu_1y_{k-1}}{2}\biggl)\biggl]^2
\biggl).
\label{eq:Uty_continuum}
\end{align}
To evaluate this integral, it is advantageous to 
slightly rewrite original Eq.~(\ref{eq:func_integral})
in the following way
\begin{align}  
  \hat{\URM}_t\left(\dfrac{\mu_1}{2}y_{n}^{2},\dfrac{\mu_1}{2}y_{0}^{2}\right)
  & = \exp
  \left( -\mu_1 t N_0 
  \right) \int 
  \fmeasure{y}
  \exp \biggl(
  -\int\limits_{0}^{t} \frac{\dRM s}{2}
  \biggl[
  \dot{y}(s)^2
  \nonumber\\
  & +\frac{\left(N_0-\frac{1}{2}\right)^2}{y^2(s)}+
  \frac{\mu_1^2y^2(s)}{4}-\left(N_0-\frac{1}{2}\right)\mu_1
  \biggl]
  \biggl) \nonumber\\
  & \times \exp
  \left(
  \frac{1}{2}\int\limits_{0}^{t} \dRM s
  \left[
  \frac{\dot{y}(s)}{y(s)}(2N_0-1)-\mu_1\dot{y}(s)y(s)
  \right]
  \right).
  \label{eq:Uty_discrete4}
\end{align}
Note that due to the functional term  $\exp
\left(-\int\limits_{0}^{t}\dRM s\,\dot{y}^2(s)/2
\right)$ this integral corresponds to a path integral representation of the Wiener process.
Here, we interpret it in the It\^o sense when the left point of the time interval is always taken in the discretized construction of the integral \citep{Gardiner}.
 In practical terms, this simply amounts to augmenting the definition of the Heaviside step function 
 with
$\theta(0) = 0$ \citep{Janssen92,Tauber2014}. 
Furthermore, It\^o calculus \citep{Gardiner} yields
 the
following relations for infinitesimal differentials and related
integral expressions
\begin{align}    
  \dRM y^2 & =\dRM s + 2 y \dRM y,
&\int\limits_{y(0)}^{y(t)}y \,\dRM y &=\frac{1}{2} 
y^2\biggl|_{y(0)}^{y(t)}-
\frac{1}{2}\int\limits_{y(0)}^{y(t)}{\dRM s}, \\
\dRM 
\ln y
& =-\frac{\dRM s}{2y^2}+\frac{\dRM y}{y},
&\int\limits_{0}^{t}\frac{\dRM y}{y} & =
\ln y 
\biggl|_{y(0)}^{y(t)}+\int\limits_{0}^{t}\frac{\dRM s}{2y^2(s)},
\end{align}
where $s$ is an independent time-like variable.
  Using these relations we rewrite the last exponential term in 
  Eq.~(\ref{eq:Uty_discrete4}) as follows
\begin{align}
\int\limits_{0}^{t} \dRM s &\left(\frac{\dot{y}(s)}{y(s)}(2N_0-1)-\mu_1\dot{y}(s)y(s)\right) =
(2N-1)  
\biggl[
\ln y
\biggl|_{y(0)}^{y(t)}
\nonumber\\
& + \int\limits_{0}^{t}\frac{\dRM s}{2y^2(s)}
\biggl]
-
\mu_1\left( \frac{1}{2} y^2\biggl|_{y(0)}^{y(t)}-\frac{t}{2}\right)
.
\end{align}
 Putting all the necessary expressions together, we arrive at
\begin{align}
\hat{\URM}_t\left(\dfrac{\mu_1}{2}y_{n}^{2},\dfrac{\mu_1}{2}y_{0}^{2}\right)
& =
\exp 
\left(
-\frac{1}{2}\mu_1 t N_0
\right)
\exp
\biggl(
\left[N_0 - \frac{1}{2}\right] \ln y \biggl|_{y(0)}^{y(t)}
\nonumber\\
& -
\mu_1 \frac{1}{4} y^2\biggl|_{y(0)}^{y(t)}
\biggl)
\int \fmeasure{ y} 
\exp
\biggl(
-\frac12\int\limits_{0}^{t} \dRM s\,
\biggl[\dot{y}^2(s) 
\nonumber\\
& -
\frac{(N_0-\frac{1}{2})(\frac{3}{2}-N_0)}{y^2(s)}
+\frac{\mu_1^2y^2(s)}{4}\biggl]
\biggl)
.
\end{align}
 This formula will be later analyzed by the approximation scheme
in Sec.~\ref{sec:calculation}.

Probabilities $\pRM_n=\pRM_n(t)$ can be compactly 
encompassed by a probability generating function $\Phi(z)$, which can be 
defined as
\begin{equation}
\Phi_t(z) \equiv \sum\limits_{n=0}^{\infty}\pRM_n(t)z^n.
 \label{eq:phi_t}
\end{equation}
Because of the normalization condition \eqref{eq:conditions}
 we immediately have $\Phi_t(1) = 1$ valid for any time $t \ge 0$. 
 Probabilities $\pRM_j(t)$ can then be derived from $\Phi_t(z)$ by taking an appropriate number of derivatives of $\Phi_t(z)$, i.e.,
\begin{equation}
  \pRM_n(t)=\dfrac{1}{n!}\left.
  \dfrac{\dRM^n\Phi_t(z)}{\dRM z^n}\right|_{z=0}.
  \label{eq:phi_derivatives}
\end{equation}
Let us observe that functions $\Phi_t(z)$ and $\Phi_0(z)$ are
related to each other through the solution (\ref{eq:formal_solution})
for the corresponding Fock states $\ket{\varphi(t)}$ and $\ket{\varphi_0}$. Utilizing relations (\ref{eq:kernel_states})
and (\ref{eq:kernel_product}) we express
  the function $\Phi_t(z)$ in terms of $\URM_t(z,\xi)$ as follows
\begin{equation}  
  \Phi_t(z)=\int\frac{\dRM\xi \dRM\xi'}{2\pi}\exp
  \left(-\iRM\xi\xi' \right)\URM_t(z,\xi)\Phi_0(\iRM\xi').
  \label{eq:phit_fun1}
\end{equation}

Further, we need a formula that relates $\hat{\URM}_t$ to a
generating function $\Phi_t(z)$. To this end, we 
 insert $z=i\psi'_n$ into Eq.~(\ref{eq:Utz_discrete2}) and get
\begin{equation}   
   \hat{\URM}_t(\psi_{n},\psi_{0})=
\int\frac{\dRM \psi'_{n}}{2\pi}\URM_t(i\psi'_{n},\psi_{0})
\exp\left(
 -\iRM\psi'_{n}\psi_{n}\right).
 \label{eq:Ut_fourier}
\end{equation}
We observe that $\hat{\URM}_t(\psi_{n},\psi_{0})$ is the Fourier transform of the kernel function 
 $\URM_t(i\psi_n',\psi_0)$. Under standard assumptions, we can readily invert this relation
\begin{equation}
  \URM_t(i\psi'_n,\psi_0)= \int
  \dRM\psi_n \, \hat{\URM}_t(\psi_n,\psi_0)\exp
  \left( \iRM\psi'_n\psi_n \right).
\label{eq:Ut_inverted}
\end{equation}
 Using this formula 
we finally express $\Phi_t(z)$ in terms of the function
$\hat{\URM}_t(\psi_{n},\psi_{0})$ 
\begin{equation}
  \Phi_t(z)=\int\frac{\dRM\xi\,
  \dRM\xi' \, \dRM\psi_n}{2\pi}\exp 
  \left( -\iRM\xi
  \xi' + \iRM \psi'_n\psi_n \right)
  \hat{\URM}_t(\psi_n,\xi)  \Phi_0(\iRM\xi').
  \label{eq:phit_fun2}
\end{equation}
To proceed, it is necessary to specify initial conditions. 
Without loss of generality, we
assume the following initial conditions
\begin{equation}
    \pRM_n (0) = \begin{cases}
      1 \quad n = 0,\\
      0 \quad n \ge 1.
    \end{cases}    
\end{equation}
For generating function (\ref{eq:phi_t}) we then simply have $\Phi_0(z)=1$. 
In Eq.~(\ref{eq:phit_fun2}) we can immediately
integrate over the variable $\xi'$ and get
\begin{align}
  \Phi_t(z) & = \int \dRM\xi\,
  \dRM\psi_n \,\delta( \xi )\hat{\URM}_t(\psi_n,\xi)
  \exp\left( z \psi_n\right) 
  \nonumber\\
  & =  \int \dRM\psi_n\, \hat{\URM}_t(\psi_n,0)\exp
  \left( z \psi_n\right).
  \label{eq:phit_fun3}
\end{align}
By the straightforward insertion of Eq.~(\ref{eq:phit_fun3}) into (\ref{eq:phi_derivatives}) the expression for
the probability $\pRM_j$ becomes
\begin{equation}
  \pRM_j(t) = 
  \frac{1}{j!}\int
  \dRM\psi_n\,\hat{\URM}_t(\psi_n,0)\psi_n^j.
\end{equation}
Note that in terms of the variable $y$ the last integral takes the form
\begin{equation}
  \pRM_j(t) =  \frac{1}{j!}\int
  \dRM y_n\,
  \hat{\URM}_t\left(\frac{\mu_1}{2}y_{n}^{2},0\right)
  \left(\frac{\mu_1}{2}y_{n}^{2}\right)^j .
\end{equation}

{\section{Approximation scheme for an evaluation of
functional integrals} \label{sec:calculation}}
Functional integrals are notoriously difficult for their exact
treatment. Besides certain special cases, e.g., quadratic Lagrangians 
\citep{gelfand1960,schulman2012}
and a few others, to gain physical information from the functional representation it is almost 
unavoidable to invoke some approximation procedure.
Here, we rely
on an elaborate numerical scheme developed by some of the present
authors in the past (see~\citep{Hnatic2023} and references therein),
 whose main points can be briefly summarized as follows.
 Once an integral representation of the time evolution 
 kernel $\URM_t$ is derived, we express the kernel through the eigenvalues and eigenvectors of the associated Hamilton operator.
 This is subsequently tackled utilizing a particular approximation scheme. 

In contrast to the formulas for a given degree of accuracy that
are useful for short time intervals \citep{risken1984}, the
method based on the eigenfunction
expansion is especially useful in the opposite asymptotic limit that of large time intervals. 
Since this
is the most relevant asymptotic region for statistical problems such as (\ref{eq:basic}), we expect this method to be profitable in this regard. For more specific details of this approximate scheme, the interested reader is referred to works \citep{Hnatic2023,ayryan2014,malyutin2018,malyutin2019,malyutin2020,ayryan2020,ayryan2021}.

Let us thus consider
the approximate calculation of the arising functional integrals from Sec.~\ref{sec:specific}
and concentrate on the calculation of the probability $\pRM_j(t)$ using the approximate calculation of the integrals  $\int \dRM\psi_n\hat{\URM}_t(\psi_n,0)\psi_n^j$. The relevant integral to be analyzed takes the form
\begin{align}
\int \dRM & \psi_n\, \hat{\URM}_t(\psi_n,0)\psi_n^j 
 =
\int \dRM y_n\,\hat{\URM}_t\left(\dfrac{\mu_1}{2}y_{n}^{2},0\right)\left(\dfrac{\mu_1}{2}y_{n}^{2}\right)^j  \nonumber\\
& = \int \dRM y_n \exp 
\left(
-\frac{1}{2}\mu_1 t N_0
\right)
\exp
\biggl[
\left(N_0 - \frac12\right)
\ln y \biggl|_{0}^{t}-
\mu_1 \frac{1}{4} y^2 \biggl|_{0}^{t}
\biggl]
\nonumber\\
& \times \left(\dfrac{\mu_1}{2}y_{n}^{2}\right)^j 
\int \fmeasure{y} 
\exp
\biggl(
-\frac{1}{2}\int\limits_{0}^{t} \dRM s \biggl(\dot{y}^2(s)-
\frac{(N_0-\frac{1}{2})(\frac{3}{2}-N_0)}{y^2(s)}
\nonumber\\
& +  \frac{\mu_1^2y^2(s)}{4}\biggl)
\biggl)
.
\label{eq:6}
\end{align}
 To approximately treat this functional integral, we
 utilize an aforementioned method based on the use of expansion
 in eigenfunctions of the Hamilton operator 
 defining the functional integral.
 From the functional integral in Eq.~(\ref{eq:6}) we observe that the Hamilton operator takes the form
\begin{equation}
H=\frac{1}{2}\left(\frac{\partial^2}{\partial
y^2}-\frac{(\frac{1}{2}-N_0)(\frac{3}{2}-N_0)}{y^2}-\frac{\mu_1^2y^2}{4}\right).
\end{equation}
 The main outcome of this approximate approach is the following representation for
  the kernel of the operator 
  $\exp \left( t H \right)$
\begin{align}
& \int 
\fmeasure{y}
\exp
\biggl(
-\int\limits_{0}^{t} \frac{\dRM s}{2}\biggl(\dot{y}^2(s)-
\frac{(N_0-\frac{1}{2})(\frac{3}{2}-N_0)}{y^2(s)}
+ 
\frac{\mu_1^2y^2(s)}{4}\biggl)
\biggl)
\nonumber\\
& \approx 
\sum\limits_{k = 0 }^{K}\exp
\left(
\lambda_k t
\right)
\phi_k(0)\phi_k(y_n),
  \label{eq:approximate_int}
\end{align}
 where  $\lambda_k$ are eigenvalues of the Hamilton operator
 $H$, and  $\phi_k = \phi_k(y)$ are corresponding eigenvectors.
 Eigenvalues $\lambda_k$ were calculated using the
 Sturm sequence method, and, on the other hand,
 eigenvectors
 $\phi_k(y)$ were obtained using the backward iteration method (more detailed exposition can be found elsewhere, e.g., in \citep{Hnatic2023,wilkinson1988}), and $(K+1)$
 is the number of eigenfunctions
 taken into account and represents a chosen
 degree of precision.
   
   Inserting the expression (\ref{eq:approximate_int}) into (\ref{eq:6}) 
   and approximating the integral over the variable $\dRM y_n$
   by the integral on the interval $[0,2\Upsilon]$ we find
   an approximate formula 
\begin{align}    
 \pRM_j(t) & \approx \frac{1}{j!}\exp
 \left(
 -\frac12\mu_1 t N_0 \
 \right)
\sum\limits_{k=0}^{K}\exp
\left(
\lambda_k t
\right)
\frac{\phi_k(0)}{y_0^{N_0 - \frac12}}
\nonumber \\
& \times  \int\limits_{0}^{2\Upsilon}
\dRM y_n\, y_n^{N_0 - \frac{1}{2}}\exp
\left(
-\frac1 4\mu_1 y_n^2
\right)
\left(\frac{\mu_1y_n^2}{2}\right)^j\phi_k(y_n) .
\label{eq:moments_approx_int}
\end{align}
Employing the trapezoidal rule \citep{krylov1972}  for an approximate calculation of the integral over $\dRM y_n$ on
 the interval $[0,2\Upsilon]$ and assuming zero boundary
conditions at points  $y_0=0$ and $y_n=2\Upsilon$, we 
finally arrive at the approximate formula for probabilities $\pRM_j(t)$ 
\begin{align}
\pRM_j(t)  & \approx 
\frac{1}{j!}\exp
\left(
 - \frac{1}{2}\mu_1 t N_0
\right)
\sum\limits_{k=0}^{K}\exp
\left(
\lambda_kt
\right)
\frac{\phi_k(0)}{y_0^{N_0 - \frac{1}{2}}} \nonumber\\
& \times  
h
\sum\limits_{m=1}^{M-1}(hm)^{N_0 - \frac{1}{2}}
\exp
\left(
 - \frac1 4\mu_1 (hm)^2
\right)
\left(\frac{\mu_1(hm)^2}{2}\right)^j\phi_k(hm),
\label{eq:moments_approx_sum}
\end{align}
 where for brevity we have introduced the abbreviation 
 $h = 2\Upsilon / M $, $M$ being the number of intervals
 into which the interval $[0,2\Upsilon]$ is divided for the approximate calculation of eigenvalues and eigenfunctions.

\begin{figure}
\begin{center}
\includegraphics[width=0.45\textwidth]{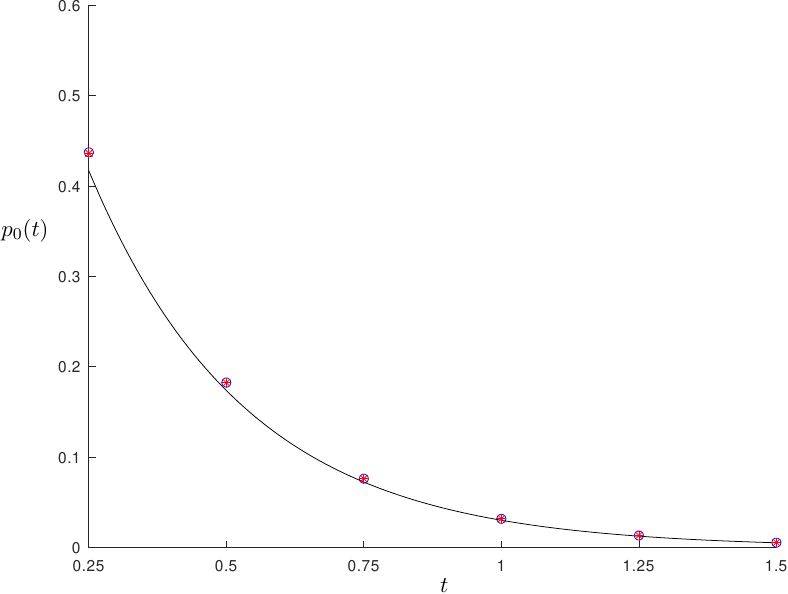}
\end{center}
\vspace*{-1\baselineskip}
\caption{Exact (solid curve) and approximate values of the probability $\pRM_0(t)$ for $\mu_1=2$, $N_0 = 7/4$, $\Upsilon=2$, $i=40$, $K=0$ (blue circles), $K=1$ (red asterisks). }
\label{fig:p0}

\bigskip

\begin{center}
\includegraphics[width=0.45\textwidth]{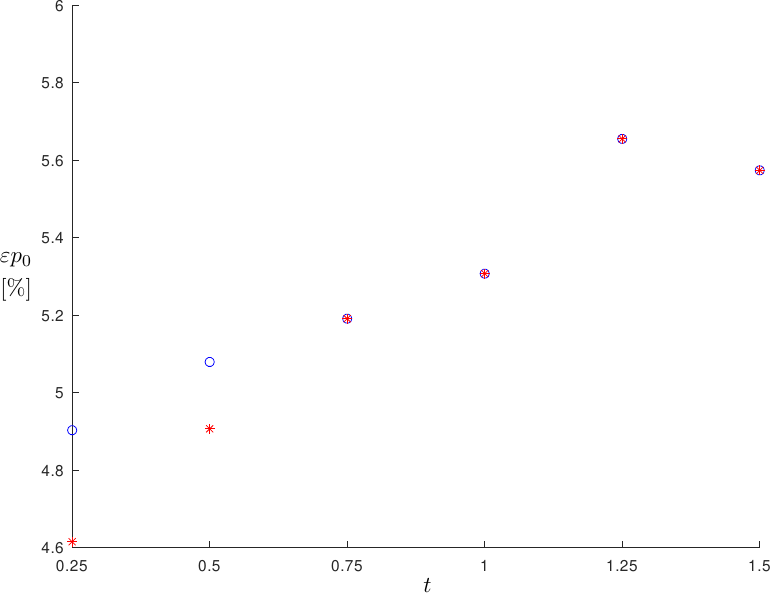}
\end{center}
\vspace*{-1\baselineskip}
\caption{Relative errors of the probabilities $\pRM_0(t)$ in \% for $K=0$ (blue circles), $K=1$ (red asterisks). }
\label{fig:relp0}
\end{figure}

\begin{figure}
\begin{center}
\includegraphics[width=0.45\textwidth]{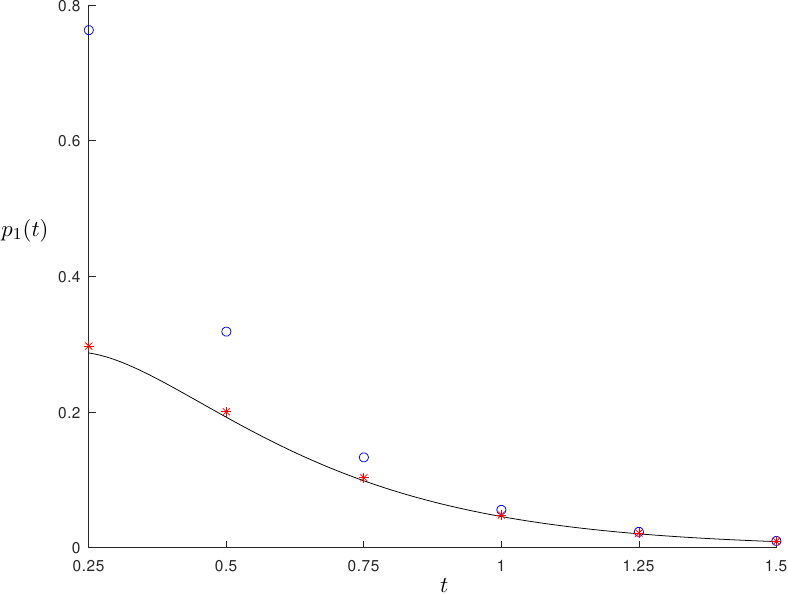}
\end{center}
\vspace*{-1\baselineskip}
\caption{ Exact (solid curve) and approximate values of the probability $\pRM_1(t)$ for  $\mu_1=2$, $N_0 = 7/4$, $\Upsilon=2$, $i=40$, $K=0$ (blue circles), $K=1$ (red asterisks).}\label{fig:p1}

\bigskip

\begin{center}
\includegraphics[width=0.45\textwidth]{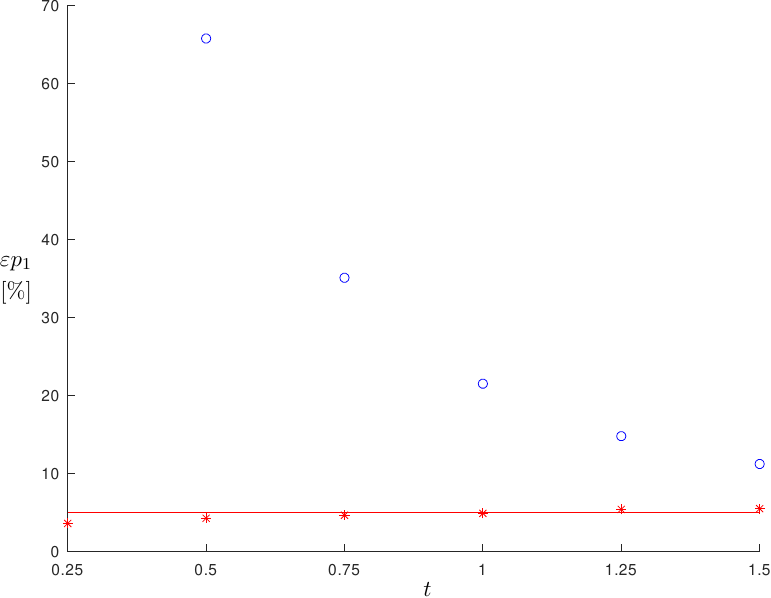}
\end{center}
\vspace*{-1\baselineskip}
\caption{Relative errors of the probabilities $\pRM_1(t)$ in \% for $K=0$ (blue circles), $K=1$ (red asterisks). }
\label{fig:relp1}
\end{figure}

 For  $\mu_1=2$, $N_0 = 7/4$, $\Upsilon = 2$, $M=40$  Figs.~\ref{fig:p0}
 and~\ref{fig:p1} show exact and approximate values for probabilities $\pRM_0(t)$ and $\pRM_1(t)$, respectively. Figs.~\ref{fig:relp0}
 and~\ref{fig:relp1} show the relative errors of the calculated probabilities (in \%). Specific values for time variable $t$ were chosen from the interval $[0.25;1.50]$. 
 Parameter $K$ that gives the number of eigenfunctions taken into account in the approximate scheme attains two values, i.e., $K = 0$ and $K = 1$, respectively. Let us stress that an already rather low value of $K = 1$ (corresponding
 to just two lowest eigenvalues) yields quantitatively satisfactory results.

Further, we can straightforwardly derive exact expressions for the probabilities $\pRM_0(t)$ and $\pRM_1(t)$ in a closed form \citep{karlin2014first}. The essential relations take the following form
\begin{align}
 \pRM_0(t) & =  
 \exp \left(-\lambda_0t\right),
 \label{eq:result_p0}
 \\
 \pRM_k(t) & =  \lambda_{k-1}
\exp\left( -\lambda_k t \right)
\int\limits_{0}^{t} \dRM s\,
\exp \left( \lambda_k s \right)
\pRM_{k-1}(s),
\label{eq:result_pk}
\end{align}
where $k \ge 1$, and parameters $\lambda_n$ have been defined previously in Eq.~(\ref{eq:model_param}).
By the repeated application of the recursion 
formula~(\ref{eq:result_pk})
we get expected results for the first several probabilities
\begin{align}
\pRM_0(t) & = \exp \left( - N_0 \mu_1 t \right), \\
\pRM_1(t) & = N_0 \exp
 \left( -N_0 \mu_1 t \right)
\left[ 1-\exp \left( -\mu_1 t\right) \right],\\
\pRM_2(t) & = \frac{N_0(N_0+1)}{2}\exp\left( -N_0 \mu_1 t \right)
\left[
  1 - \exp\left(-\mu_1 t \right)
\right]^2.
\end{align}
We can even infer
an explicit solution for integer $k\ge 1$ in the form
\begin{equation}
    \pRM_k(t) = \binom{N_0 + k - 1}{k}
    \exp\left( -N_0 \mu_1 t \right)
\left[
  1 - \exp\left(-\mu_1 t \right)
\right]^{k}.
\end{equation}
Further by using the well-known Newton's binomial theorem \citep{markushevich1965} 
\begin{equation}
   (1+x)^{-n} = \sum_{k=0}^\infty 
   \binom{n+k-1}{k}(-x)^k, 
\end{equation}
which holds for $x\in (-1,1)$, 
it is possible to find a closed expression for a given moment $M_k$ introduced in Eq.~(\ref{eq:moments}).
For instance, for the average $M_1(t) = \overline{n}$ we find
\begin{equation}
  M_1(t) = N_0  \left[
  \exp(\mu_1 t) - 1
  \right],
\end{equation}
whereas for the variance $\sigma^2= \overline{n^2} - \overline{n}^2$ (mean squared deviation) we obtain
\begin{equation}
    \sigma^2 = M_2 - M_1^2 = 
    N_0 \exp\left(2\mu_1 t\right)\left[
    1 - \exp\left(-\mu_1 t\right)
    \right].
\end{equation}
{\section{Conclusion} \label{sec:conclusion}}
In this paper, we have analyzed a particular version
of a stochastic one-step process employing the formalism of functional integrals.
We have given a detailed exposition of theoretical techniques
that allow us to recast the underlying master equation 
into a functional integral representation suitable for an approximate treatment. 
Based on our previous results we have proposed a method for approximate calculation of arising functional integrals.
  Using this method, we have obtained approximate values of the probabilities of being in the $n$-th state ($n=0,1,\ldots$) for
  the pure birth process.
We believe that the proposed technique can be formulated and applied to similar classical problems tractable by the operator approach and we hope to extend the formalism in future studies.

\section*{Acknowledgements}
The work was supported by VEGA grant 
No. 1/0297/25 of the Ministry of Education, Science, 
Research and Sport of the Slovak Republic. 

\bibliographystyle{elsarticle-harv} 
\bibliography{mybib}

\end{document}